\documentclass[a4paper,useAMS,usenatbib]{mn2e}
\usepackage[totalwidth=480pt, totalheight=680pt]{geometry}
\usepackage{graphicx}
\usepackage{hyperref}

\def\gr{{$\gamma$-ray~}}
\def\ergs{{erg~cm$^{-2}$~s$^{-1}$~}}

\newcommand{\lsim}{{\lower.5ex\hbox{$\; \buildrel < \over \sim \;$}}}
\newcommand{\gsim}{{\lower.5ex\hbox{$\; \buildrel > \over \sim \;$}}}
\def\nup{$\nu_{\rm peak}^S$}
\def\avvovm{{$\langle V/V_{\rm m} \rangle$}}

\newcommand{\fermi}{{\it Fermi} }
\newcommand{\planck}{{\it Planck}}
\newcommand{\swift}{{\it Swift} }

\title[A simplified view of blazars]{A simplified view of blazars:
clearing the fog around long-standing selection effects}
 \author[P. Giommi et al.]{P. Giommi$^{1}$\thanks{E-mail:
paolo.giommi@asdc.asi.it}, P. Padovani$^{2}$, G. Polenta$^{1,3}$, S. 
Turriziani$^{1}$, V. D'Elia$^{1,3}$,  
\newauthor
S. Piranomonte$^{3}$\\
$^{1}$ASI Science Data Center, c/o ESRIN, via G. Galilei, 00044 Frascati, Italy \\
$^{2}$European Southern Observatory, Karl-Schwarzschild-Str. 2,
D-85748 Garching bei M\"unchen, Germany\\
$^{3}$INAF-Osservatorio Astronomico di Roma, via Frascati 33, I-00040
Monteporzio Catone, Italy}

\begin{document}

\date{Accepted 2011 October 20. Received 2011 October 4; in original form 2011 July 13}

\pagerange{\pageref{firstpage}--\pageref{lastpage}} \pubyear{2011}

\maketitle

\label{firstpage}

\begin{abstract}
  We propose a scenario where blazars are classified as flat-spectrum radio
  quasars (FSRQs), BL Lacs, low synchrotron, or high synchrotron peaked
  objects according to a varying mix of the Doppler boosted radiation from
  the jet, the emission from the accretion disk, the broad line region, and
  the light from the host galaxy. In this framework the peak energy of the
  synchrotron power (\nup) in blazars is independent of source type and of
  radio luminosity.  We test this new approach, which builds upon unified
  schemes, using extensive Monte Carlo simulations and show that it can
  provide simple answers to a number of long-standing issues including,
  amongst others, the different cosmological evolution of BL Lacs selected
  in the radio and X-ray bands, the larger \nup~values observed in BL Lacs,
  the fact that high synchrotron peaked blazars are always of the BL Lac
  type, and the existence of FSRQ/BL Lac transition objects.  Objects so
  far classified as BL Lacs on the basis of their {\it observed} weak, or
  undetectable, emission lines are of two physically different classes:
  intrinsically weak lined objects, more common in X-ray selected samples,
  and heavily diluted broad lined sources, more frequent in radio selected
  samples, which explains some of the confusion in the literature. We also
  show that strong selection effects are the main cause of the diversity
  observed in radio and X-ray samples, and that the correlation between
  luminosity and \nup, that led to the proposal of the ``blazar sequence'',
  is also a selection effect arising from the comparison of shallow radio
  and X-ray surveys, and to the fact that high \nup~-~high radio power
  objects have never been considered because their redshift is not
  measurable.
\end{abstract}



\section{Introduction}\label{intro}

Once considered rare sources, blazars, a type of radio loud active galactic
nuclei (AGN) pointing their jets in the direction of the observer
\citep[see e.g. ][]{bla78,UP95}, are now being detected in increasingly
larger numbers. Recent results from the {\it Wilkinson Microwave Anisotropy
  Probe} (WMAP), the \planck~and \fermi satellites have established
that blazars are the most common type of extragalactic sources found at
microwave and \gr energies \citep{GiommiWMAP07,fermi1lac, planckERCSC}.  So
far about 3,000 blazars are known \citep{mas09,mas11}, but their number is
steadily growing thanks to the \fermi \citep{fermi1fgl,fermi2fgl}, the
optical Sloan Digital Sky Survey \citep[SDSS:][]{Plotkin10}, and the
\planck~\citep{planckERCSC} surveys. Some faint blazars are also being
detected as serendipitous sources in {\it Swift}-XRT images \citep{tur10,tur11}.

While all blazars share the same property of emitting variable, non-thermal
radiation across the entire electromagnetic spectrum, they also display
diversity. Namely, they come in two main subclasses, whose major difference
is in their optical properties: 1) Flat Spectrum Radio Quasars (FSRQs),
which show strong, broad emission lines in their optical spectrum, just
like radio quiet QSOs; and 2) BL Lacs, which are instead characterized by
an optical spectrum, which at most shows weak emission lines, sometimes
displays absorption features, and in some cases can be completely
featureless. Historically, the separation between BL Lacs and FSRQs has
been made at the (rather arbitrary) rest-frame equivalent width (EW) of
5~\AA~\citep[e.g.][]{stickel1991,sto91}. However, no evidence for a bimodal
distribution in the EW of the broad lines of radio quasars has ever been
found and, on the contrary, \cite{sca97} pointed out that radio-selected BL
Lacs were, from the point of view of the emission line properties, very
similar to FSRQs but with a stronger continuum.  Most BL Lacs selected in
the X-ray band, on the other hand, had very weak, if any, emission lines,
and \cite{sto91}, when studying the properties of the X-ray selected {\it
  Einstein} Medium Sensitivity Survey (EMSS) sample, had to introduce
another criterion to identify BL Lacs, this time to separate them from
galaxies. This was based on the Ca H\&K break, a stellar absorption feature
typically found in the spectra of elliptical galaxies. Given that its value
in non-active ellipticals is $\sim 50\%$, \cite{sto91} chose a maximum
value of $25\%$ to ensure the presence of a substantial non-thermal
continuum superposed to the host galaxy spectrum. This was later revised to
$40\%$ \citep{mar96,lan02}.

\begin{figure*}
\includegraphics[height=7.8cm,angle=-90]{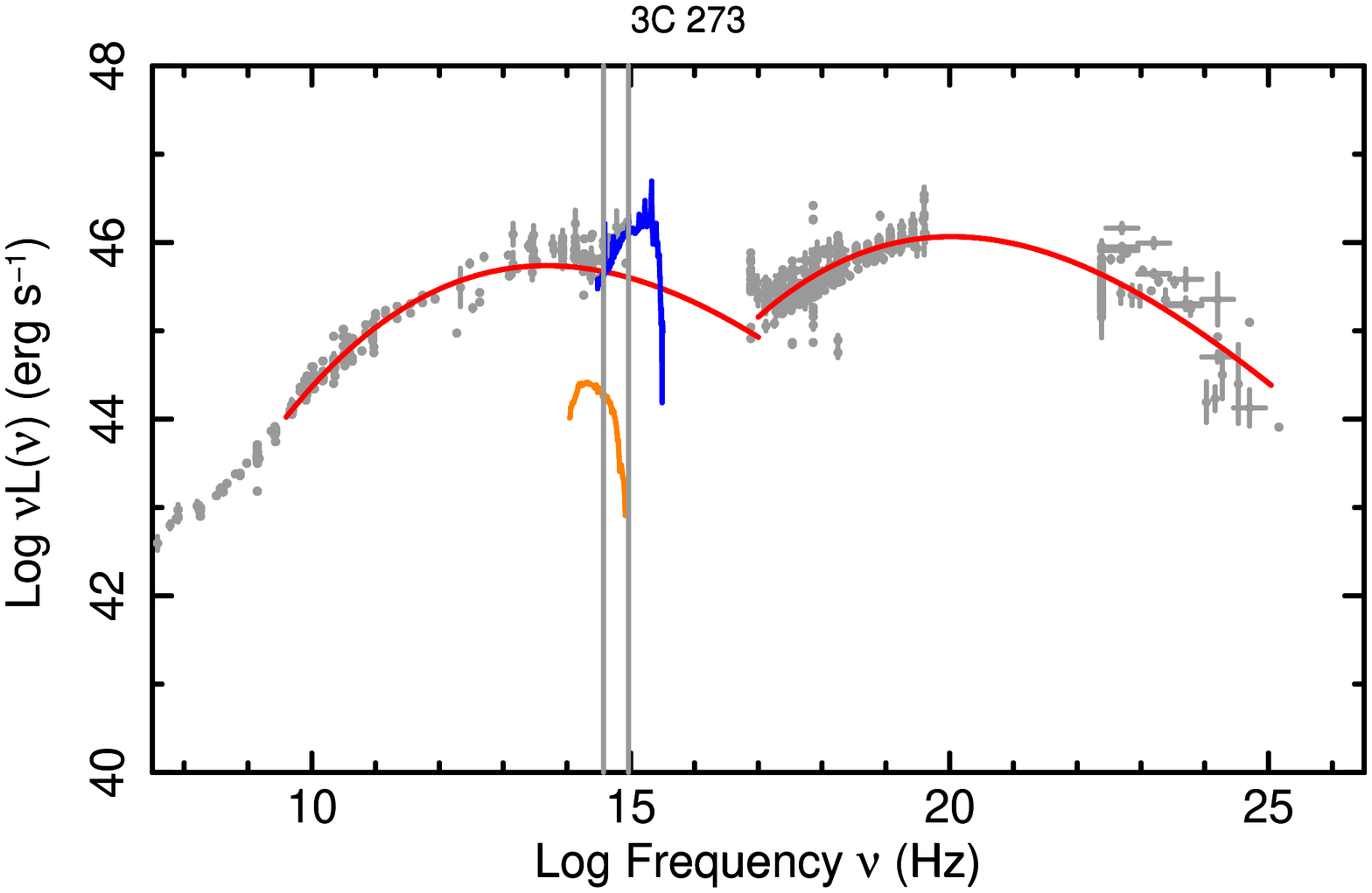}
\hspace{0.3cm}
\includegraphics[height=7.8cm,angle=-90]{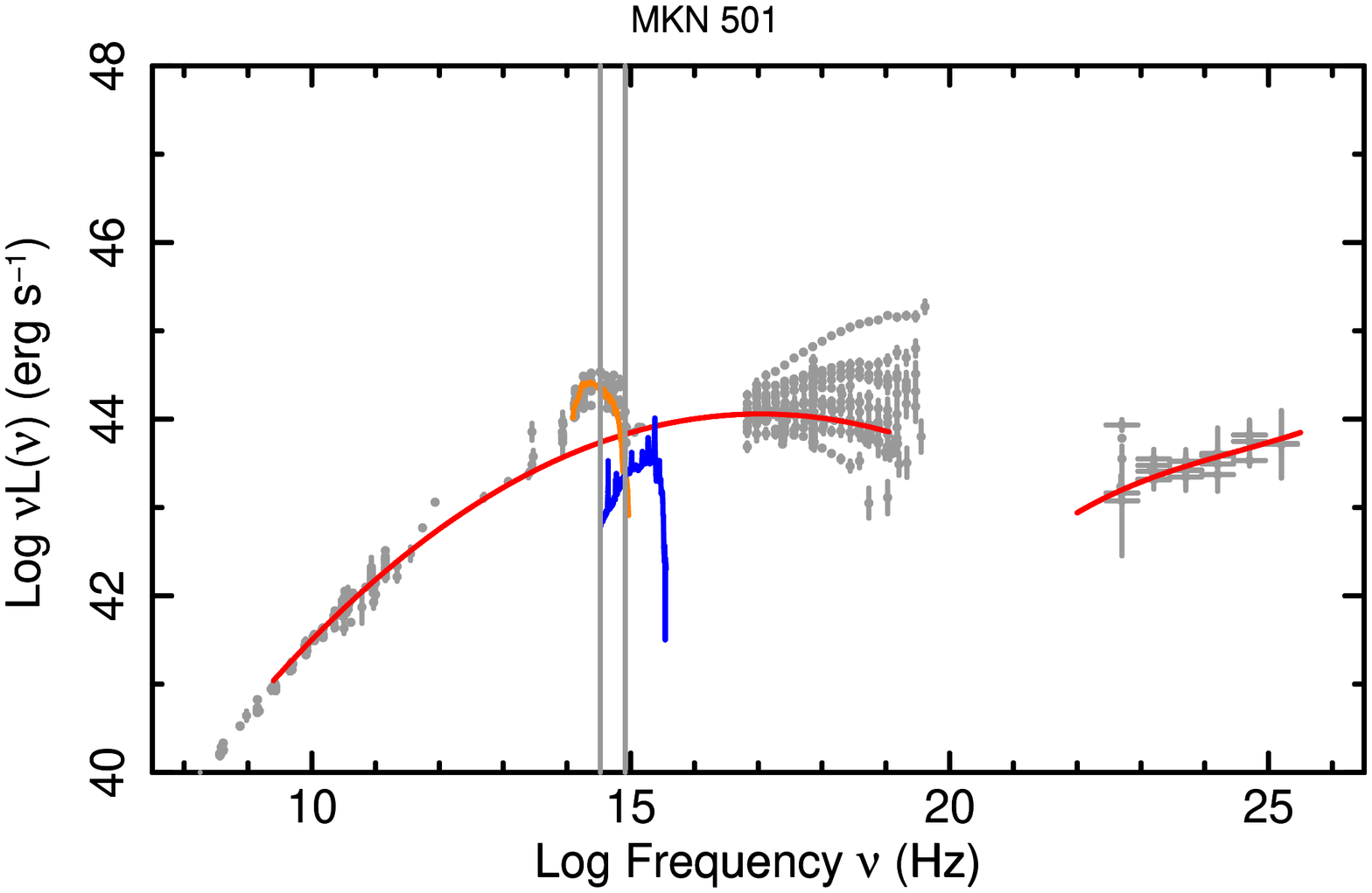}\\
\vspace{2.5cm}
\includegraphics[height=7.8cm,angle=-90]{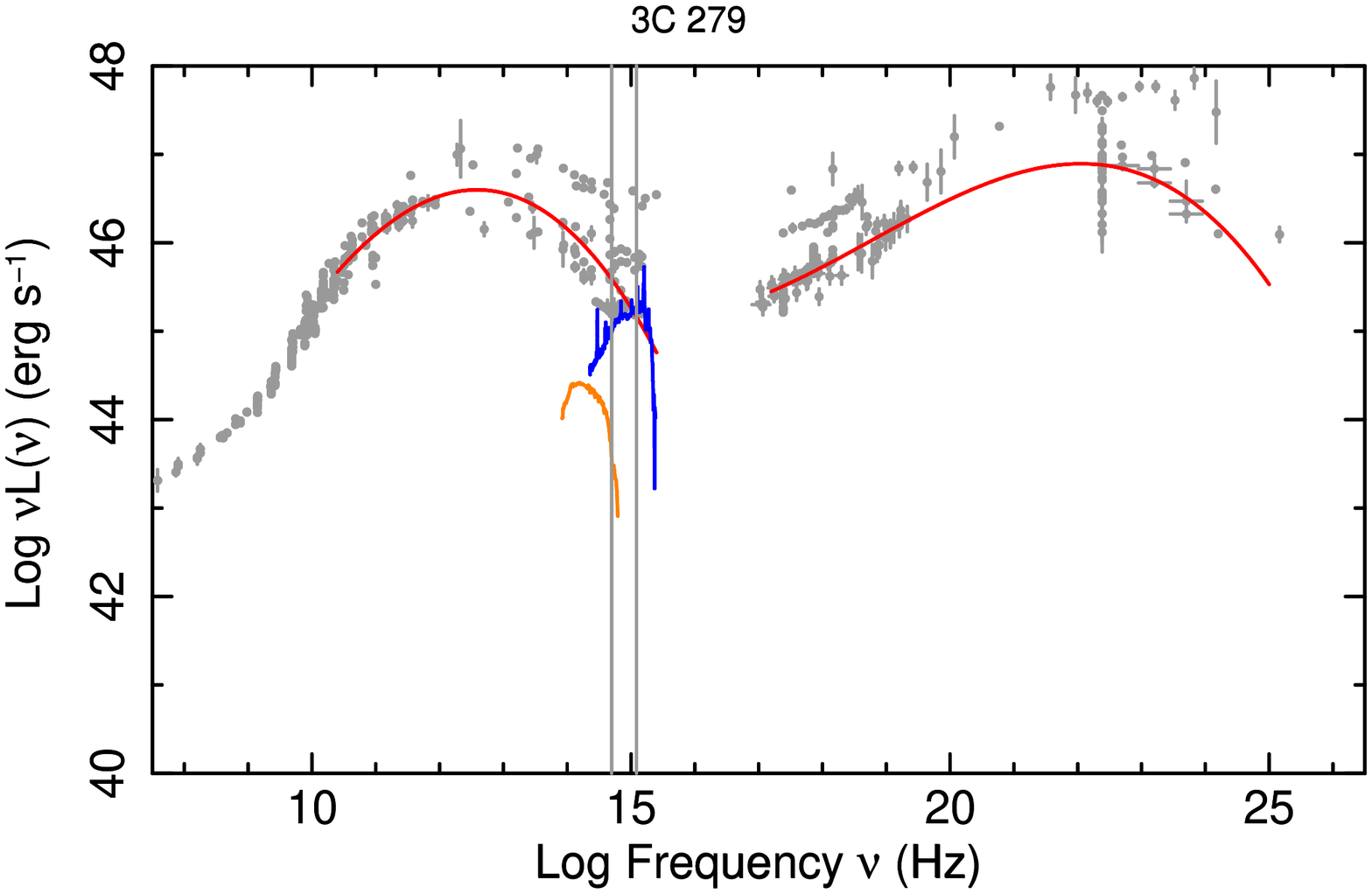}
\hspace{0.3cm}
\includegraphics[height=7.8cm,angle=-90]{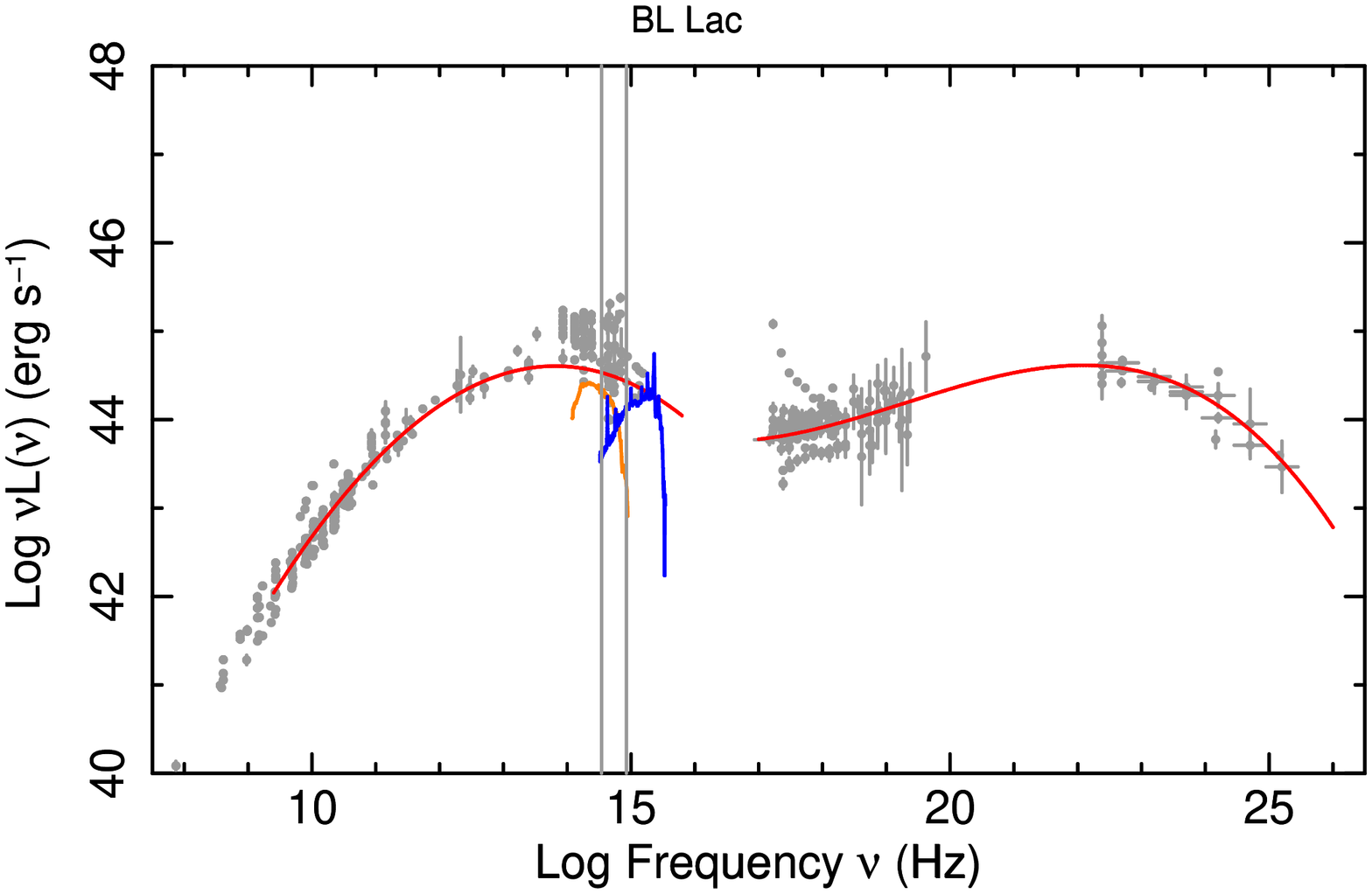}
\vspace{2.0cm}
\caption{The SEDs of four representative blazars: two FSRQs, 3C 273 and 
3C 279, and two BL Lacs, MKN 501 and BL Lac. The lines in color denote the three main 
  components of blazars SEDs, namely non-thermal radiation from the 
  jet (red), emission from the disk and from the broad line region  
 represented by the composite QSO optical spectrum of \protect \cite{van01} 
  (blue), and light from the host galaxy, represented by the giant elliptical 
  template of \protect \cite{man01} (orange). 
 The two vertical lines indicate the optical observing window ($3800 - 8000$ \AA). See text for details.}
\label{fig:sed}
\end{figure*}

Blazar classification depends then on the details of their appearance in
the optical band where they emit a mix of three types of radiation: 1) a
non-thermal, jet-related, component; 2) thermal radiation coming from the
accretion onto the supermassive black hole and from the broad line region
(at least in most radio-selected sources); 3) light from the host (giant
elliptical) galaxy. Figure \ref{fig:sed} represents these three components
as red, blue and orange lines, overlaid to the spectral energy distribution
(SED) of four well-known blazars \citep[from][]{GiommiPlanck}. The strong
non-thermal radiation, the only one that spans the entire electromagnetic
spectrum, is composed of two basic parts forming two broad humps, the 
low-energy one attributed to synchrotron radiation, and the high-energy one,
usually thought to be due to inverse Compton radiation \citep[see
e.g.][]{abdosed}. The peak of the synchrotron hump (\nup) can occur at
different frequencies, ranging from about $\sim 10^{12.5}$~Hz to over
$10^{18}$~Hz (see e.g.  the cases of 3C 273 or 3C 279 and MKN 501 in
Fig. \ref{fig:sed}) reflecting the maximum energy at which particles can be
accelerated \citep[e.g.][]{GiommiPlanck}. Blazars where \nup~is lower than
$10^{14}$~Hz in their rest frame are called Low Synchrotron Peaked (LSP)
sources, while those where $10^{14}$~Hz $<$ \nup~$< 10^{15}$~Hz, and \nup~
$> 10^{15}$~Hz are called Intermediate and High Synchrotron Peaked (ISP and
HSP) sources respectively \citep{abdosed}. This definition extends the
original division of BL Lacs into LBL and HBL sources first introduced by
\cite{padgio95}.

The large \nup~disparity between LSP and HSP blazars (up to five orders of
magnitude) is the cause of very large differences between the intensity of
the radiation emitted in different energy bands. For instance, for the same
radio flux an HSP source can be a factor of a 100 brighter in the optical band,
or even a factor of a 1,000 brighter in the X-ray band, than an LSP blazar. This
induces very strong selection effects in blazar samples discovered in
different bands and led to some confusion when comparing, for
example, the first radio and X-ray-selected BL Lac samples in the early
1990's.

We feel that all the above factors, which play an important role in blazar
classification, have not been properly taken into account so far. The
purpose of this paper is then to propose a new hypothesis to explain the
existence of apparently different properties of the blazar subclasses,
which solves at once many of the open issues of blazar research.

Throughout this paper we use a $\Lambda$CDM cosmology with $H_0 = 70$ km
s$^{-1}$ Mpc$^{-1}$, $\Omega_m = 0.27$ and $\Omega_\Lambda = 0.73$
\citep{kom11}.

\section{Current status: two types of blazar populations with widely different properties}

The two main blazar subclasses have many differences, which include:

\begin {enumerate}

\item {\it different optical spectra} (by definition). There are, however, a
  number of BL Lac - FSRQ transition objects, which include even BL
  Lacertae itself, the prototype of the class, which displays at times
  moderately strong, broad lines \citep[e.g.][]{ver96,cap10,ghis11} and 3C
  279, a well-studied FSRQ, which can appear nearly featureless in a bright
  state \citep{pia99};


\item {\it different extended radio powers}. Most BL Lacs have extended radio
  powers and morphologies consistent with those of Fanaroff-Riley (FR) type
  I, while basically all FSRQs are FR II-like \citep[][and references
  therein]{UP95}. However, despite the difficulty of classifying the radio
  morphology of sources with their jets forming a small angle with respect
  to the line of sight, some radio-selected BL Lacs are known to posses an
  FR II-like structure \citep[e.g.][]{kol92,rec01};

\begin{figure}
\includegraphics[height=8.cm, angle=-90]{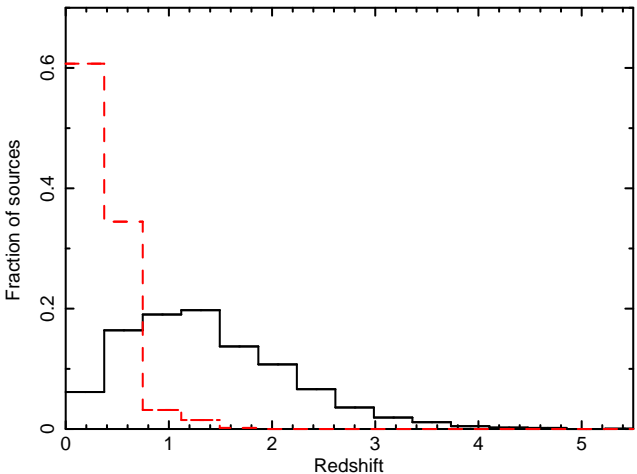}
\vspace{0.8cm}
\caption{The redshift distribution of FSRQs (1676 objects, solid line) and
  BL Lacs (537 objects, dashed line) in the third edition of the BZCAT
  catalogue \citep{mas09,mas11}. The two distributions are obviously
  different at the $>99.99$ level, with means of 1.4 and 0.36
  respectively. About 400 BL Lacs with no redshift determination in the
  BZCAT catalogue have been excluded.}
\label{fig:histoz}
\end{figure}

\item {\it very different redshift distributions.} FSRQs, similarly to radio
  quiet QSOs, are typically found at redshifts $\sim 1-2$, and up to $\sim
  5.5$, while BL Lacs are usually much closer with very few cases at $z
  \gsim 0.6$. This is graphically shown in Fig. \ref{fig:histoz}, which
  plots the redshift distribution of all the FSRQs and BL Lacs included in
  the third edition of BZCAT \citep{mas09,mas11}, the largest compilation of
  blazars currently available. It must be stressed, however, that a large
  fraction of BL Lacs \citep[$\sim 43$\% in BZCAT and $> 50 - 60$\% of the
  BL Lacs in the Fermi 1 and 2 year AGN catalogs:][]{fermi1lac,fermi2lac}
  have no measured redshift, due to the lack of any detectable feature in
  their optical spectrum, despite the use of 10-m class optical
  telescopes for the spectroscopy identification campaign
  \citep{shaw09,shaw10};
   
\item {\it different cosmological evolutions.} Detailed analyses of both radio
  selected and X-ray selected samples have led to the conclusion that the
  cosmological evolution of the two blazar subclasses are very different,
  with FSRQs evolving strongly (again similarly to radio quiet QSOs) and BL
  Lacs evolving at a similar, or perhaps lower rate, in the radio band, or
  even showing no or negative evolution in the X-ray band \citep[e.g.][and
  references  therein]{stickel1991,giommi_sedent_I,rec00,beck03,pad07,GiommiWMAP09};

\item {\it widely different mix of FSRQs and BL Lacs in radio and X-ray selected
  samples}, with the latter typically including a much larger fraction of BL
  Lacs than the former. In fact, while only $\sim 15\%$ of WMAP5 blazars
  are BL Lacs (Section \ref{ingredients}), this fraction is instead $\sim
  70\%$ in the EMSS complete sample, which includes 41 BL Lacs and 15 FSRQs
  \citep{rec00,pad03};

\item {\it widely different distributions of the synchrotron peak energy \nup.} 
  The rest-frame \nup~distribution of FSRQs is strongly peaked at low
  energies ($ \langle$ \nup $\rangle=10^{13.1\pm0.1}$ Hz) and never reaches
  very high values (\nup $\lsim 10^{14.5}$ Hz) independently of the
  selection method \citep{GiommiPlanck}, while the \nup~ distribution of BL
  Lacs is shifted to higher values by at least one order of magnitude. It
  can also reach values as high as \nup $\gsim 10^{18}$ Hz and its shape
  varies strongly depending on the selection band \citep[that is radio,
  X-ray or $\gamma$-ray:][]{abdosed,GiommiPlanck}.
  
\end {enumerate}

Some of these differences have been explained by so-called unified schemes,
which posit that BL Lacs and FSRQs are simply FR I and FR II radio galaxies
with their jets forming a small angle with respect to the line of sight
\citep{UP95}. (Radio galaxies would then be the ``parent'' population of
blazars). Due to relativistic beaming this has enormous effects on their
apparent emitted power and luminosity functions (LFs) and can explain their
different extended radio powers and, partly, their cosmological
evolutions. However, unified schemes per se cannot account for transition
objects, and the different evolution of radio and X-ray selected BL Lacs
and \nup~distributions. Our new hypothesis builds upon unified schemes by
adding dilution and strong selection effects as new, vital components.

\section{The EW of FSRQs and radio quiet QSOs}\label{sec:EW}

The existence of BL Lac - FSRQ transition objects and the recent finding
that unresolved radio quasars (i.e. FSRQs) appear to be redder than radio quiet AGN
\citep[][]{kim11} suggest that the strong non-thermal jet emission in
blazars may have a significant impact on the shape of their blue bump and
on the EW of their broad lines. In particular, given this extra continuum
component, the EWs of broad-lined blazars (FSRQs) should be systematically
lower than those of radio-quiet AGN.

To test this, we extracted EW data for samples of radio-quiet
QSOs and blazars from the SDSS DR-7 spectral database. In order to have a
sample that is not excessively large but still representative of the entire
data set, we considered only radio-quiet QSOs with $28^{\circ} \leq b_{\rm
  II} \leq30^{\circ}$, thus limiting the sample size to about 650
sources. To obtain the EW of a sizable sample of blazars we considered all
FSRQs in BZCAT with 1.4 GHz flux $> 300$ mJy (to simulate a radio
flux-limited sample) and in the WMAP5 catalogue. For each object in the
three samples we retrieved, from the SDSS on-line system, the line
measurements parameters (e.g. EW, $\chi^2_{\nu}$ of the fit etc.)  of the
strongest emission lines in quasar spectra, that is Ly$\alpha$, C~IV,
C~III, Mg~II, H$\beta$, H$\alpha$.
Details on the fitting algorithm can be found at the SDSS
website\footnote{http://www.sdss.org/dr7/algorithms/speclinefits.html}.  To
avoid problematic cases, for our analysis we considered only SDSS fits to
emission lines with $\chi^2_{\nu} \le 2.5$.

\begin{figure}
\includegraphics[height=8.cm,angle=-90]{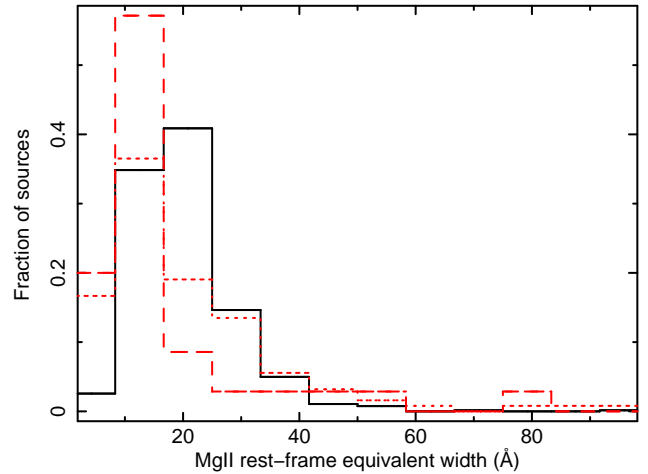}
\vspace{0.8cm}
\caption{The distribution of the Mg~II rest-frame EW for a sample of 663
  radio quiet quasars (solid histogram) compared to the distributions of
  the Mg~II EW of 35 WMAP5 and 126 FSRQs in BZCAT (dashed and dotted
  histograms respectively) with SDSS-DR7 spectrum and $\chi^2_{\nu} \le
  2.5$. The blazar EW distribution is significantly shifted towards lower
  values compared to radio quiet QSOs. See text for details.}
\label{fig:EWRLRQ}
\end{figure}

Figure \ref{fig:EWRLRQ} shows the distribution of the EW of the MgII
emission line, for which we have better statistics, for the sample of radio
quiet SDSS QSOs (663 objects), BZCAT (126 objects) and WMAP5 (35 objects)
FSRQs. The EW of both blazar samples is smaller than that of radio quiet
QSOs, with median values respectively equal to 16.4 (BZCAT), 14 (WMAP5),
and 18.4 \AA~(radio-quiet QSOs). A Kolmogorov-Smirnov (KS) test gives a
probability $< 0.1 \%$ that both FSRQ samples have the same distribution as
that of radio-quiet AGN. Moreover, we found that this dilution increases
further (i.e. the median EW gets smaller) if we make a cut at higher radio
fluxes in the BZCAT FSRQs. Similar results were obtained for the other
emission lines. This demonstrates that the broad lines of blazars are
significantly diluted compared to those of radio quiet QSOs. The mean
absolute magnitudes of the blazar and radio-quiet samples agree within
$\sim 0.2$ magnitudes, which means that the difference cannot be due to the
well-known anti-correlation between EW and absolute magnitude, the
so-called ``Baldwin effect'' \citep{bal77}.

We note that a number of papers have stated that the optical spectra of
radio-quiet and radio-loud QSOs are not significantly different
\citep[e.g.][and references therein]{fra93}. However, the radio-loud
samples on which this conclusion was based were still optically selected
and required the presence of a UV excess and/or strong emission lines.
Therefore, these radio-loud samples were biased towards objects with strong
blue bumps.

\section{A new, much simpler, scenario for blazars}

The considerations made in the previous sections lead us to propose a new,
very simple scenario where the observed blazar optical spectrum is the
result of a combination of an intrinsic EW distribution and the effects of
three components: a non-thermal, jet related one, a thermal one due to the
accretion disk, and emission from the host galaxy. Different mixes of these
components determine the appearance of the optical spectrum and therefore 
the classification of sources in FSRQs (dominated by strong lines), BL Lacs
(with diluted, weak lines, if a standard accretion disk is present), and
radio-galaxies (where the host galaxy swamps both the thermal and
non-thermal nuclear emission present in blazars). The other novel component
is a single LF whose evolution depends on radio
power. This idea was tested through extensive Monte Carlo simulations, as
described below.

\subsection{Simulation ingredients}\label{ingredients}

Our simulations include the following ingredients,
which we kept as simple as possible and tied as much as possible to
observational data:

\begin{enumerate}

\item{\bf Luminosity function} We derive the LF and evolution of blazars at
  41 GHz from the Wilkinson Microwave Anisotropy Probe (WMAP5) sample
  \citep{wri09}. Given that the mean blazar radio - sub-mm spectral index
  is $\sim 0$ \citep{GiommiWMAP09} this is practically equivalent to a
  radio-selected sample. We extend on the work of \cite{GiommiWMAP09},
  which was based on WMAP3, and define a flux-limited sample of high
  Galactic latitude sources ($f_{41GHz} \ge 0.9$ Jy, $|b_{\rm II}| >
  15^{\circ}$) including 161 FSRQs, 29 BL Lacs, and 10 blazars of unknown
  type. By applying a maximum likelihood technique to the WMAP5 blazars
  \citep[see, e.g.][for details]{pad11} we obtain, together with the
  evolution discussed below, a best-fit local LF $\Phi(P) \propto P^{-3}$
  (in units of Gpc$^{-3}$ P$^{-1}$) between $1.9 \times 10^{24}$ and $4.2
  \times 10^{27}$ W/Hz, which we assume in our radio simulations.
  
\item{\bf Cosmological evolution} Powerful ($P_{\rm r} \ga 10^{26}$ W/Hz)
  radio sources are known to display a strong, positive evolution at
  moderately low redshifts followed by a decline at higher redshifts
  \citep[e.g.][and references therein]{wa05}. We parametrize this behaviour
  with a simple model of the type $P(z) = (1+z)^{k+\beta z}$, which allows
  for a maximum in the luminosity evolution followed by a decline.  This
  was first suggested by \cite{wa08} and applied by \cite{Aje09} to a
  sample of {\it Swift}/BAT blazars. A maximum likelihood technique applied
  to the WMAP5 blazar sample allows us to derive $k = 7.3$ and $\beta =
  -1.5$ in the $0 - 3.4$ redshift range (which implies a peak at $z \sim
  1.85$), which we assume in our simulations. The case of pure luminosity
  evolution ($\beta = 0$) is excluded with very high significance ($P >
  99.99\%$). Lower luminosity ($P_{\rm r} \la 10^{26}$ W/Hz), mostly FR I
  radio sources are known to display a much weaker cosmological evolution,
  which reaches $\approx$ zero at $P_{\rm r} < 10^{25}$ W/Hz
  \citep[e.g.][and references therein]{gen10}. We took this into account by
  using the radio LFs of BL Lacs and FSRQs derived from those of FR Is and
  FR IIs and based on the beaming model of \cite{UP95}, which agree well
  with those of recent blazar samples \citep{pad07,GiommiWMAP09}. We then
  used the fraction of beamed FR I blazars in bins of radio power to
  simulate the fraction of non-evolving radio sources as a function of
  power. This fraction is equal to 1 for $P_{\rm r} \le 5 \times 10^{24}$
  W/Hz, decreases monotonically with power, and reaches 0 for $P_{\rm r}
  \ge 5 \times 10^{27}$ W/Hz.
  
\item{\bf Non-thermal component}  
To represent the non-thermal/jet component, we 
assume a simple homogeneous synchrotron self-Compton model \citep[SSC, see, e.g.][and references therein]{tra09}
with relativistic electrons distributed as a power law at low energies and as a log-parabola at high energies \citep{mas04,mas06}.
This model represents well the synchrotron part of the observed SEDs, which always extends at least to the optical band where the classification of blazars as FSRQs 
or BL Lacs occurs. 
As for the inverse Compton emission, which can be important in the soft X-ray band, we set the Compton dominance so as to reproduce the observed  $f_{\rm x}/f_{\rm r}$ in FSRQs. 
This is sufficient for our purposes, since, complicating the emission with additional components, like thermal emission
from accretion or inverse Compton on an external field of photons, would only modify somewhat the amount of observed soft X-rays in LSP blazars but 
would not  change the composition of the samples, nor alter any of our conclusions.
  
 The Lorentz factors of the electrons radiating at the peak of the
  synchrotron SED component ($\gamma_{\rm peak}$) are distributed as shown
  in Fig. \ref{fig:gammae}. The range of $\gamma_{\rm peak}$ ($\sim 2.5 -
  4.5$) is that expected for typical parameters of the SSC model as shown
  in Fig. 36 of \cite{abdosed}. The particular shape of the distribution
  was chosen so as to empirically reproduce the observed \nup~distributions
  in radio and X-ray selected samples of blazars. The skewness to lower
  values is similar to that of the $f_{\rm x}/f_{\rm r}$ (a proxy for \nup)
  distribution adopted by \cite{padgio95} to unify X-ray selected and
  radio-selected BL Lacs.  As regards the Doppler factor, we assumed a mean
  value of 15, which was chosen to be consistent with the mean superluminal
speed $\beta_{\rm app} \sim 12$ obtained by \cite{lis09}, and with the typical Lorentz factor $\Gamma \sim 15$ derived by
\cite{hov09} (since for the angle that maximizes the apparent velocity $\delta \sim \beta_{\rm app} \sim \Gamma$).
     
\begin{figure}
\includegraphics[height=8.cm,angle=-90]{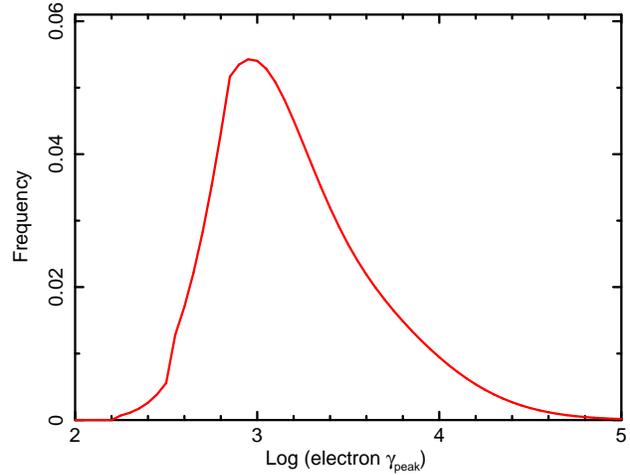}
\vspace{0.7cm}
\caption{The distribution of the Lorentz factors of the electrons radiating
  at the peak of the synchrotron SED used for the simulation, which also assumes 
  a magnetic field of B=0.15 Gauss and a gaussian distribution of Doppler factors
  with $\langle\delta\rangle$=15.}
\label{fig:gammae}
\end{figure}

\item{\bf Accretion Disk and broad emission lines}

  We use the quasar spectral template of \cite{van01}, which has been built
  using an homogeneous dataset of over 2,200 SDSS spectra \citep[see Fig. 4
  of][]{GiommiPlanck}. A standard accretion disk is likely to be present only in
  so-called ``high excitation'' radio galaxies (HERGs), while it
  appears not to be there, or be less efficient, in low-excitation ones
  (LERGs). Almost all FR Is are LERGs, while most FR IIs are HERGs,
  although there is a population of FR II LERGs as well. \cite{chi99} and
  \cite{do04} have in fact suggested that the obscuring torus, which is
  required by AGN unification schemes, is absent in FR Is, based on the
  high optical core detection rate and low X-ray intrinsic absorption,
  respectively. This would mean no accretion disk as well, otherwise one
  would see broad lines in their optical spectra, while, apart from a
  handful of objects (e.g. 3C 120), basically no FR
  I displays broad optical lines. \cite{ev06} have shown that the accretion
  flow luminosities of FR IIs are typically several orders of magnitude
  higher than those of FR Is. It then looks like LERGs, which means all FR
  Is and some FR IIs, either do not possess an accretion disk, or if the
  disk is present is much less efficient than in FR IIs (i.e. of the 
  Advection Dominated Accretion Flow [ADAF] type). 
  We have then associated the presence of a standard accretion disk only
  with beamed FR II sources and assumed that all those with an FR I parent
  (the non-evolving sources) have no disk (but see Section \ref{discussion}
  for an alternative scenario).
  
\item{\bf Equivalent width distributions} The intrinsic (before dilution)
  distributions of the EW of the broad lines (Ly$\alpha$, C~IV, C~III,
  Mg~II, H$\beta$, H$\alpha$) have been assumed to be those of the radio
  quiet QSOs included in the SDSS database described in Section
  \ref{sec:EW}.  We assumed Gaussian distributions characterized by the
  measured means ($\langle$EW${_{{\rm H}\alpha}}$$\rangle$= 200 \AA,
  $\langle$EW${_{{\rm H}\beta}}$$\rangle$= 23 \AA,
  $\langle$EW${_{{\rm Mg-II}}}$$\rangle$= 18 \AA,
  $\langle$EW${_{{\rm C-III}}}$$\rangle$= 16 \AA,
  $\langle$EW${_{{\rm C-IV}}}$$\rangle$= 20 \AA,
  $\langle$EW${_{{\rm Ly}\alpha}}$$\rangle$= 47 \AA)  and dispersions for the various lines. 
 
\item{\bf The disk to jet power ratio} The disk and jet components in
  blazars are known to be correlated and possibly of the same order of
  magnitude, although there are uncertainties associated with estimating
  them \citep[see, e.g.][]{del03,ghis11}. We are interested in the somewhat
  simpler question of determining how the luminosity of the accretion disk
  (blue bump intensity at 5000 \AA) scales with radio power (at 5 GHz). The
  relevant data were derived by using the very large amount of
  multi-frequency information included in public databases and the tools
  that are now available to analyze SEDs \citep{SEDtool}. 
  Figure \ref{fig:sed} gives some examples of representative
  objects. The amount of thermal flux in each FSRQ was estimated by  matching the composite optical QSO spectrum of \cite{van01} to the SED data in the blazar rest-frame. 
  We have done that through a careful visual inspection of each SED and by adjusting the intensity of the composite QSO spectrum until it overlapped well to the observed emission. The use of this manual 
  approach was necessary as the heterogeneity of the available data and flux variability does not allow the implementation of a robust automatic procedure.  Fig. \ref{fig:sed}  gives examples of the matching of the
  composite QSO spectrum (blue line) to the data for the case of 3C273 or 3C279.
  In those objects where the available optical/UV data are limited to a magnitude in one or two colors, we matched the composite QSO spectrum to the flux level corresponding to the
  available magnitude(s), taking into account of redshift. 
 
  An upper limit was instead estimated for BL Lac objects by placing the composite QSO
  spectrum in the SED at an intensity such that the optical lines would not
  be detectable in the optical spectrum (typically a factor ten below the observed flux. Fig. \ref{fig:sed} illustrates the case of MKN501). This was done for a large number
  of blazars selected in four surveys, two radio-selected (Deep X-ray Radio
  Blazar Sample [DXRBS] and WMAP5) and two X-ray selected (EMSS and
  {\it Swift}-BAT).
The results are plotted in Fig. \ref{fig:arovsrlum} where we can see that
$\alpha_{r-BlueBump}$ (the slope between the 5 GHz luminosity and the blue
bump intensity at 5000~\AA~and defined by $L_{\rm disk} = L_{\rm r}
(\nu_{\rm 5000\AA}/\nu_{5GHz})^{-\alpha_{\rm r-BlueBump}}$) correlates with
radio luminosity, although with a large scatter. We interpret this as the
result of relativistic beaming, in the sense that the larger the radio
luminosity (and therefore the beaming amplification) the larger the ratio
between non-thermal and thermal radiation.

We adopted a simple linear relationship between the two
variables ($\alpha_{r-BlueBump}$ = 0.04*log(L$_{\rm radio}) 
-0.39$, see Fig. \ref{fig:arovsrlum}) and assumed a Gaussian distribution 
around it with a dispersion of 0.1. This has been derived by fitting the WMAP5
and EMSS points in the Figure (see below). We did not take into account the
lower limits on $\alpha_{r-BlueBump}$ since the large majority of them are at
low radio power ($\lsim 10^{26}$ W/Hz) and in our scenario these sources
do not have a standard accretion disk.

\begin{figure}
\includegraphics[height=8.cm,angle=-90]{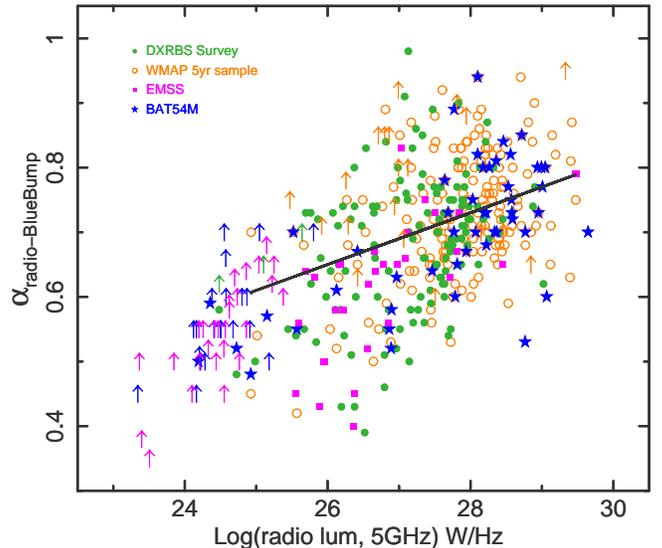}
\vspace{0.9cm}
\caption{$\alpha_{r-BlueBump}$, the slope between the radio luminosity at 5
 GHz (non-thermal boosted component) and the optical luminosity at 5000
 \AA~due to the blue bump (thermal non-boosted component estimated from
 the blazars' SEDs as described in the text), is plotted against radio
 luminosity for a large number of blazars from several radio and X-ray
 selected samples.  
 The solid line is the linear relationship used for our simulations (see text for details).
  }
\label{fig:arovsrlum}
\end{figure}

\item{\bf Host galaxy}

 Following \cite{scarpa2000} and \cite{urry00} we have assumed that the
 host galaxy of blazars is a giant elliptical with fixed absolute magnitude of
  $M_{\rm R} = -22.9$ as estimated by \cite{sba05} 
 using the same $\Lambda$CDM cosmology adopted in this paper. 
 These authors have shown that the dispersion around this value is less 
 than 1 magnitude and therefore such a giant elliptical galaxy can be 
 considered a standard candle. For the spectral shape we used the 
 galaxy template of \cite{man01}, who derived it combining the data from 
 28 local elliptical galaxies observed in the wavelength range 
 $0.12 - 2.4~\mu$m.

Figure \ref{fig:sed} shows this template superposed to the SED of four well-known blazars.

\end{enumerate}

\subsection{Simulation steps}

Our Monte Carlo simulations consist of the sequential execution of the
following steps:

\begin{enumerate}

\item draw a value for the radio luminosity and redshift of a simulated
  blazar based on the luminosity function and evolution described above;
\item draw a value of the Lorentz factor of the electron radiating at the
  peak of the synchrotron power ($\gamma_{\rm peak}$) from the distribution
  shown in Fig. \ref{fig:gammae}, which is assumed to be independent of
  blazar luminosity;
\item Calculate the peak of the synchrotron power in the source rest frame. In our simulations we assume 
a simple SSC model, and therefore $\nu_{peak}=3.2\times10^6\gamma_{\rm peak}^2 B \delta$.
The value of $\gamma_{\rm peak}$ is derived in step ii), the magnetic field is fixed to B = 0.15 Gauss, 
and the Doppler factor $\delta$ is randomly drawn from a gaussian distribution with $\langle \delta \rangle =15$ 
and $\sigma$= 2.
\item calculate the observed radio flux density (from the radio luminosity
  and redshift from step i)) and the non-thermal emission in the optical and
  X-ray bands under the assumption that the spectral shape of the observed emission is a log parabola around \nup\
  and that the low energy part of the SED (cm and mm wavelengths) is a power law, as is typically seen in blazars (see e.g. Fig. \ref{fig:sed}
  for some representative examples, or \cite{GiommiPlanck} for a much larger sample of blazar SEDs);
\item accept the source if its flux in the band under consideration (radio
  or X-ray) is above the flux limit chosen for the simulated survey;
\item add an accretion (blue bump) component as described above (only for
  beamed FR II sources), re-scaling the SDSS quasar template to this value;
\item draw a value of the equivalent width of Ly$\alpha$, C~IV, C~III,
  Mg~II, H$\beta$, H$\alpha$ starting from the EW distribution observed in
  the SDSS radio quiet QSOs;
\item add the optical light of the host galaxy assuming a standard giant
  elliptical observed in blazars;
\item calculate the total optical light and the observed equivalent width
  of all the broad lines considered by taking into account
  the dilution due to the non-thermal and host galaxy optical light;
\item classify the source as an FSRQ if the rest-frame EW of at least one
  of the broad lines that enter the optical band in the observer frame
  (which we assume to cover the $3,800 - 8,000$ \AA~range) is $>
  5$~\AA. Otherwise, the object is classified as a BL Lac, unless the host
  galaxy dominates the optical light causing the Ca H\&K break to be larger
  than 0.4 \citep{mar96,lan02}, in which case the source is classified as a
  radio galaxy. A BL Lac whose maximum EW is $< 2$ \AA,\, or the non-thermal light is at least a factor 10 
  larger then that of the host galaxy \citep{piranomonte07}, is deemed to have a
  redshift which cannot be typically measured.

\end{enumerate}

It is important to stress that the scope of our simulations is {\it not} to
reproduce {\it all} the observational details. While that could be possible
in theory, in practice it would require a large number of parameters and
some speculations. Our approach is instead to keep the number of
assumptions to a minimum, with the aim of allowing us to obtain robust,
almost model-independent conclusions. Our main results are in fact
independent of the simulation details (see Section \ref{sec:stability}).

\subsection{Simulations of radio and X-ray surveys}

We simulated a radio flux density limited survey with $f \ge 0.9$ Jy, to
match the WMAP5 sample, 
and an X-ray flux limited survey down to $5\times
10^{-13}$ \ergs in the $0.3 - 3.5$ keV band, in order to be able to compare
it with the EMSS. To ensure good statistics each simulation run included 10,000 sources. 
In the X-ray case, since radio powers reach lower values than in the radio case, we
extrapolated the radio LF down to $1.9 \times 10^{23}$ W/Hz assuming the same
slope.
    
\section{Comparing simulations and real data}

In this section we make a detailed comparison between the results of our
simulations and the observational data in terms of fractions of BL Lacs and
FSRQs, redshift and \nup~distributions, and cosmological evolution using
the $V/V_{\rm m}$ test \citep{sch68}, where $V$ is the volume out to the source 
and $V_{\rm m}$ is the the volume at the distance where the object would be 
at the flux limit. We also show where our simulated 
sources end up on the plot used by \cite{fossati98} when they first described
the ``blazar sequence''. 
 
\subsection{ Radio flux density limited survey}\label{radio_survey}

Table \ref{tab:radiosim} summarizes our main results by giving the number
of sources per class, their mean redshift, and \avvovm. The number
in parenthesis refers to the BL Lacs with measurable redshift, to which the
mean redshift and \avvovm~pertain. About 3/4 of our sources are
classified as FSRQs, with the fraction of BL Lacs being $\sim 19.8\%$ of
blazars, which is consistent with the value of $15.3^{+3.7}_{-3.0} \%$ in
the WMAP5 sample \citep[where the $1\sigma$ errors are based on binomial
statistics:][]{geh86}. A small fraction ($5 \%$) of the simulated blazars
are classified as radio galaxies. These are bona-fide blazars misclassified
because their non-thermal radiation is not strong enough to dilute the host
galaxy component.

\begin{figure}
\includegraphics[height=8.cm,angle=-90]{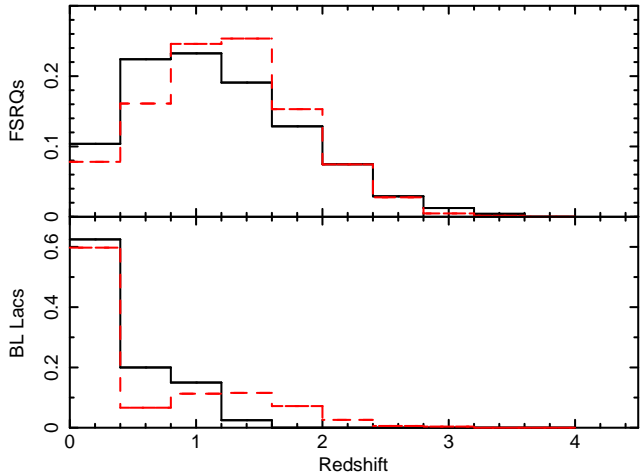}
\vspace{0.8cm}
\caption{Top panel: the redshift distribution of the WMAP5 FSRQs (solid
  histogram) compared to that of FSRQs in a simulation of a radio survey
  (dashed histogram). Bottom panel: the redshift distribution of the WMAP5
  BL Lacs (solid histogram) compared to that of BL Lacs in a simulation of
  a radio survey (dashed histogram)}
 \label{fig:wmap_red}
\end{figure}

The mean redshift for our simulated FSRQs is in good agreement with the
WMAP5 value of $1.13$, while for BL Lacs this is slightly larger than the
WMAP5 mean ($0.55$). 
 
Fig.\ref{fig:wmap_red} shows the overall good agreement between our simulated
redshift distributions (where we have only included sources with a
measurable redshift) and the observed ones.

$63\%$ of our BL Lacs have a redshift determination, in excellent agreement
with the WMAP5 value of $69^{+27}_{-20} \%$. 
$79\%$ of the BL Lacs ($68\%$ of those with redshift) have a standard accretion 
disk and are therefore broad-lined but are classified as BL Lacs
only because their observable emission lines are swamped by the non-thermal
continuum.

\begin{table}
\caption{Results from a simulation of a radio flux density limited survey (0.9 Jy)}
\begin{tabular}{llcc}
 Source type & Number of & &  \\
   & sources &  $\langle z \rangle$ &  $\langle V/V_{\rm m} \rangle$  \\
\hline
FSRQs      &    ~7,587        &  1.24 & 0.64  \\
BL Lacs     &     ~1,879 (1,191)   &  0.87 & 0.60 \\
Radio galaxies    &     ~~~534     &  0.04 & 0.48   \\
\hline
Total     &    10,000       &  1.13 & 0.63  \\
\hline
\end{tabular}
\label{tab:radiosim}
\end{table}
 
As regards \avvovm, our simulated mean values agree with the observed
ones of $0.62\pm0.02$ and $0.63\pm0.05$ for WMAP5 FSRQs and BL Lacs
respectively \citep[cf. also the value of $0.60\pm0.05$ for the 1 Jy BL Lac
sample:][]{stickel1991}.

\begin{figure}
\includegraphics[height=8.cm,angle=-90]{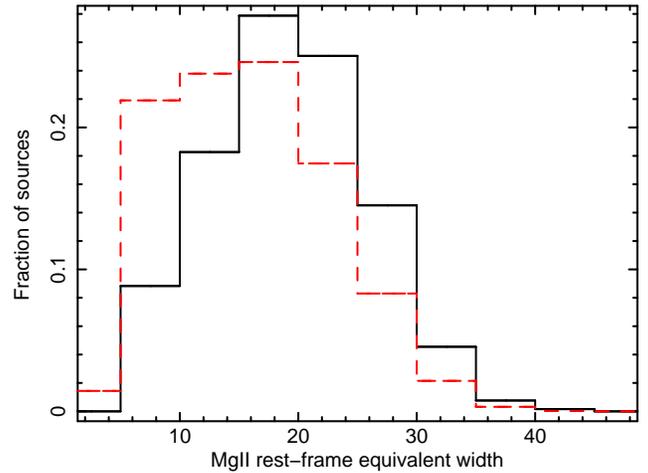}
\vspace{0.7cm}
\caption{The Mg II EW distribution used as input for our simulation (solid
  line) and the output, diluted distribution for FSRQs (dashed line) in the
  simulated radio survey.}
\label{fig:EWsim}
\end{figure}

Fig. \ref{fig:EWsim} shows the EW distribution used as input for our
simulation and the resulting one for FSRQs after the dilution due to the
non-thermal and host galaxy components. The significant shift to lower
values is clearly seen. This is similar to the comparison between 
real FSRQs and radio quiet QSOs shown in Fig. \ref{fig:EWRLRQ}.

\begin{figure}
\includegraphics[height=8.cm,angle=-90]{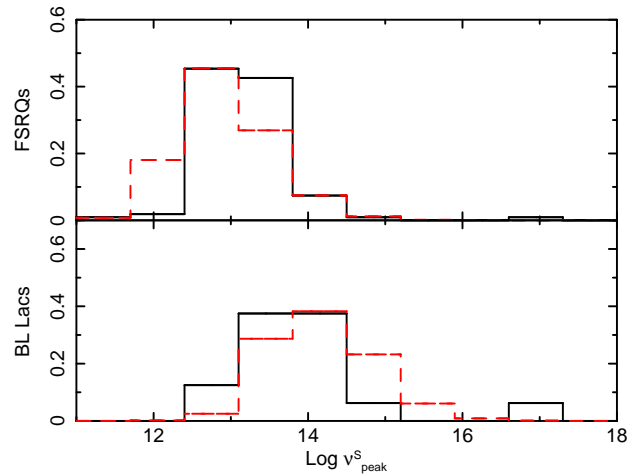}
\vspace{0.5cm}
\caption{Top panel: the \nup~distribution of radio selected FSRQs taken
  from the work of \citet{GiommiPlanck} (solid histogram) compared to that
  of FSRQs in a simulation of a radio survey (dashed histogram). Bottom
  panel: the \nup~distribution of the BL Lacs in the radio sample of
  \citet{GiommiPlanck} (solid histogram) compared to that of BL Lacs in a
  simulation of a radio survey (dashed histogram).}
 \label{fig:radio_nupeak}
\end{figure}

Fig. \ref{fig:radio_nupeak} compares the distributions of \nup, the
synchrotron peak energy, of sources classified as FSRQs and BL Lacs in our
simulation with those of blazars included the radio sample of
\citet{GiommiPlanck}, which is the sample with the best determination of
\nup~values currently available. The agreement is clearly quite good and
reproduces well the fact that BL Lacs tend to have \nup~values
significantly higher than FSRQs.

We note that, although our input parameters were partly based on the WMAP5
sample, it is important to stress that, starting from a {\it single} LF and
evolution, plus a fraction of non-evolving sources, we are able to
reproduce quite well the main properties of the two blazar subclasses
including those not part of the input (like relative fractions and
percentage of BL Lacs with redshift).

\begin{figure}
\includegraphics[height=8.cm,angle=-90]{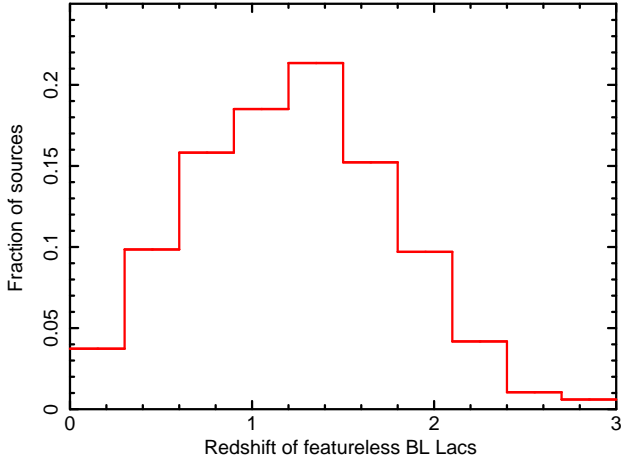}
\vspace{0.8cm}
\caption{The redshift distribution of the BL Lacs that show a featureless spectrum in our simulation of a radio flux density limited survey, and that in a real survey would have no redshift determination.}
 \label{fig:zdistOfnoredshiftSources}
\end{figure}

\subsection{X-ray flux limited survey}

Table \ref{tab:xsim} summarizes our main results. About 2/3 of our blazars
are classified as BL Lacs, which is consistent with the value of
$73^{+19}_{-15} \%$ in the EMSS sample. As in the radio case, a small
fraction ($15\%$) of the simulated blazars are misclassified as radio
galaxies.

The mean redshifts for our simulated FSRQs and BL Lacs are in reasonable
agreement with the EMSS blazar sample values $\sim 1$ and $\sim 0.37$. We
note that, unlike the WMAP5 sample, the EMSS sample is relatively small (56
sources) and therefore a detailed comparison is hampered by the small
number statistics.

$79\%$ of our BL Lacs have a redshift determination, in good agreement with
the EMSS value of $93^{+26}_{-21} \%$. Although we assumed that all
non-evolving sources do not have a standard accretion disk, $30\%$ of the
BL Lacs possess one and are classified as BL Lacs only because their
emission lines are swamped by the non-thermal continuum. The smaller
fraction of X-ray selected BL Lacs with disks in our simulations, as
compared to radio-selected ones, is in accordance with the fact that fewer
EMSS BL Lacs have emission lines clearly detectable in their optical
spectra than, for example, 1 Jy BL Lacs \citep[see,
e.g.][]{rec00,rec01,stickel1993}.

\begin{table}
\caption{Results from a simulation of an X-ray  flux limited survey ($5\times 10^{-13}$ \ergs)}
\begin{tabular}{llcc}
 Source type & Number of &  &    \\
   & sources &   $\langle z \rangle$ &  $\langle V/V_{\rm m} \rangle$  \\
\hline
FSRQs      &    ~2,836       &  1.23 & 0.65  \\
BL Lacs (all)   &     ~5,622 (4,460)    &  0.36 &  0.51 \\
BL Lacs  (log \nup~$> 16.5$) &     ~~~927 (895)      &  0.33 &  0.45 \\
BL Lacs (log \nup~$> 17$) &     ~~~185 (177)      &  0.34 &  0.34 \\
Radio galaxies    &    ~1,542    &  0.04 & 0.48   \\
\hline
Total     &    10,000        &  0.58 &   0.55  \\
\hline

\end{tabular}
\label{tab:xsim}
\end{table}

As regards $V/V_{\rm m}$, our simulated mean values agree reasonably well
with the EMSS ones of $0.67\pm0.08$ and $0.42\pm0.05$ for FSRQs and BL Lacs
respectively, derived using the samples described in \cite{pad03}.

\begin{figure}
\includegraphics[height=8.cm,angle=-90]{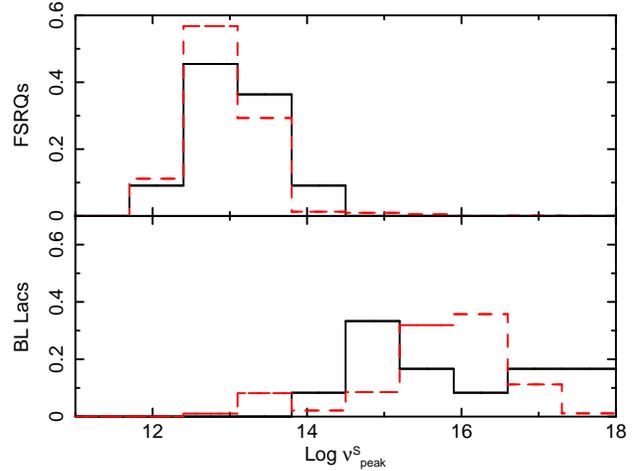}
\vspace{0.5cm}
\caption{Top panel: the \nup~distribution of the X-ray selected FSRQs in
  \citet{GiommiPlanck} (solid histogram) compared to that of FSRQs in a
  simulation of an X-ray survey (dashed histogram). Bottom panel: the
  \nup~distribution of the BL Lacs in the X-ray flux limited sample of
  \citet{GiommiPlanck} (solid histogram) compared to that of BL Lacs in a
  simulation of an X-ray survey (dashed histogram).}
\label{fig:x_nupeak}
\end{figure}

Fig. \ref{fig:x_nupeak} compares the distributions of \nup~of FSRQs and BL
Lacs in our simulation with those of blazars belonging to the soft X-ray sample
of \citet{GiommiPlanck}, which includes \planck, \swift and \fermi observed
blazars and it is therefore probably the sample with the best determination
of \nup~values currently available. The agreement is clearly good
reproducing well the fact that BL Lacs have much higher \nup~values than
FSRQs. 





\subsection{A blazar sequence?}\label{sec:seq}

\begin{figure*}
\includegraphics[height=14.cm,angle=-90]{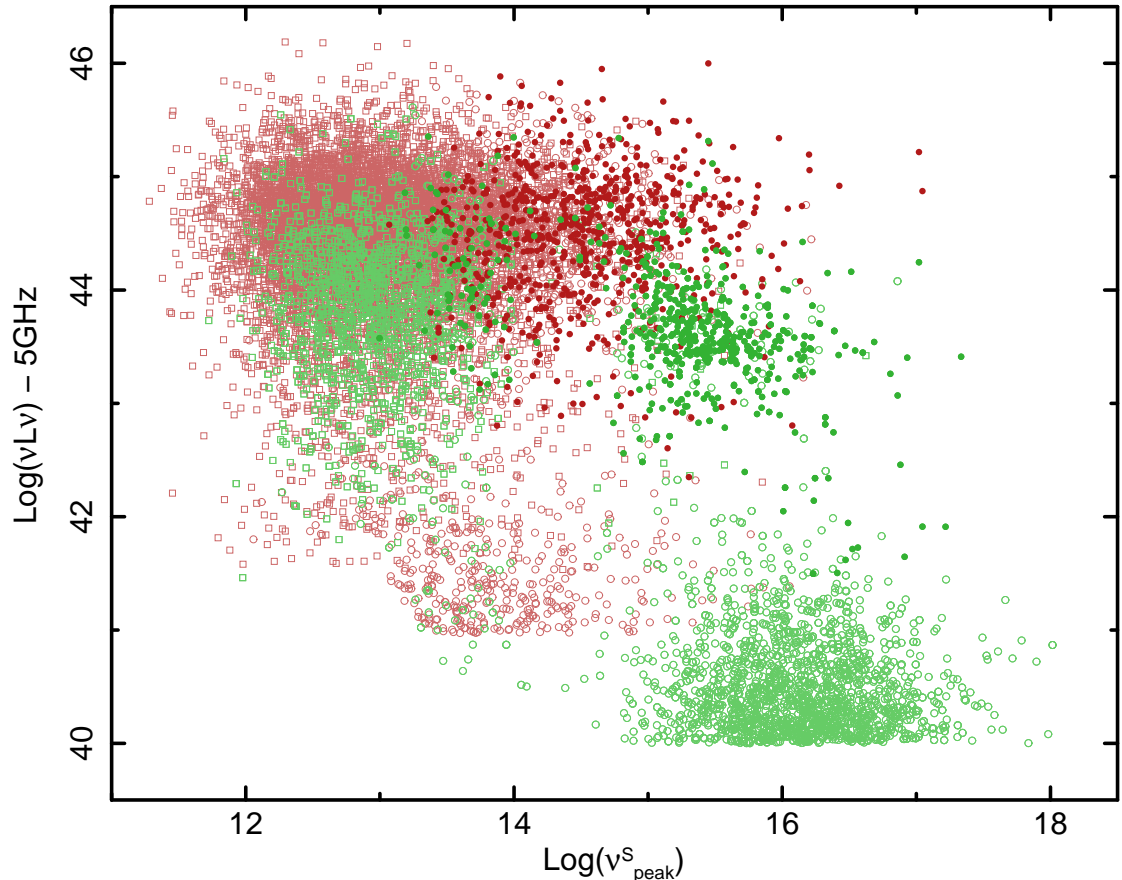}
\vspace{0.9cm}
\caption{The FSRQs (open squares) and BL Lacs (circles) of our radio flux
  density limited ($f_{\rm r} > 0.9$ Jy, red symbols) and X-ray flux
  limited ($f_{\rm x} > 5\times 10^{-13}$\ergs , green symbols) simulated
  samples, plotted in the log($\nu$)-log($\nu$L$_{\nu}$) plane, the portion
  of the parameter space used by \citet{fossati98} to show the existence of
  the ``blazar sequence'' by comparing radio ($f_{\rm r} > 1-2$ Jy) and
  X-ray selected ($f_{\rm x} \gsim 10^{-12}$\ergs) complete samples.
  Filled circles represent BL Lacs with very weak lines (EW $< 2~$\AA) or
  completely featureless which, in a real survey would most probably not
  have a measured redshift, and therefore would not appear in the plot.
  Note that almost all of these objects occupy the top right part of the
  diagram. Blazars in this area are high-luminosity HSP sources, that is
  objects that are ``forbidden'' in the blazar sequence scenario.}
\label{fig:sequence}
\end{figure*}

The existence of a strong anti-correlation between bolometric luminosity and \nup, known as the ``blazar sequence"  has been the subject of intense discussions since its first proposal by \cite{fossati98} and \cite{ghis98} \citep[e.g.,][]{giommi_sedent_I,padovani03,caccianiga04,nieppola06,padovani07,ghisellinitavecchio08,nieppola08,GiommiPlanck}. In this section we use our simulations to comment on the existence of such a  
 sequence.

Figure \ref{fig:sequence} shows our radio and X-ray selected simulated
blazars in the log(\nup) -- log($\nu$L$_{\nu}$(5 GHz)) plane. This reproduces the
famous plot used by \citet{fossati98} to propose the existence of the
blazar sequence based on the correlation shown in this plane
by FSRQs and BL Lacs discovered in shallow radio surveys (2 and 1 Jy
samples) and BL Lacs found in the X-ray flux limited {\it Einstein} slew
survey ($f_{\rm x} \gsim 10^{-12}$\ergs). Indeed, considered together
the simulated radio and X-ray selected blazars display a broad correlation
with radio selected FSRQs and BL Lacs (red open squares and open circles)
mostly filling the top left and central part of the diagram and 
X-ray selected BL Lacs (green open circles) mostly confined to the
lower right corner of the plot. This particular positioning of
the points (bright FSRQs of the LSP type vs. fainter HSP BL Lacs) is not due
to any intrinsic correlation between luminosity and \nup~but is instead
the result of a selection effect resulting from the fact that bright radio
sources are mostly drawn from the high end of the blazar luminosity
function, while BL Lacs in X-ray flux limited samples are mostly high
\nup~sources (intrinsically rare) drawn from the low end of the luminosity
function where the source density is largest. The most important difference
between Fig. \ref{fig:sequence} and the diagram of \citet{fossati98} is in
the high-luminosity - high \nup~part, where most of the radio and X-ray selected
objects with no redshift (red and green filled points) are located. These
sources could not be plotted by  \cite{fossati98} since the luminosity of
blazars without redshift cannot be estimated. This left the top right part of the 
 \cite{fossati98}  diagram empty, thus contributing to make the displayed data look like a power sequence. 

\subsection{Assessing the stability of our results}\label{sec:stability}

To assess the dependence of our results on the adopted LF and evolution, we
have made the following two checks: first, we varied their input values by
$1 \sigma$ adopting $\Phi(P) \propto P^{-2.85}$, $k=7.0$, and $\beta =
-1.4$ (luminosity peak at $z \sim 1.9$) and $\Phi(P) \propto P^{-3.1}$,
$k=7.7$, and $\beta = -1.7$ (luminosity peak at $z \sim 1.75$); second, we
used as an alternative LF the sum of the BL Lac and FSRQ LFs based on the
beaming model of \cite{UP95} (converted to $H_0 = 70$ km s$^{-1}$
Mpc$^{-1}$) assuming, for consistency with the way they were derived, a
pure luminosity evolution of the type $P(z) = P(0) {\rm exp} [T(z)/\tau]$,
where $T(z)$ is the look-back time. A value of $\tau = 0.33$, which is
consistent with the evolution of DXRBS FSRQs and BL Lacs combined
\citep[based on the samples in][]{pad07}, was also assumed. 

We have also checked the dependence of our results on the assumed value of the
Doppler factor, running the simulation with values of  $\langle\delta\rangle$ ranging from 
5 to 20, and we have also considered the case of a dependence of $\delta$ on radio power as 
suggested by some authors \citep[e.g.][]{hov09}, although \cite{lis09}, who
studied a complete sample, have not been able to confirm such a
correlation. 

In all cases our main results, that is the prevalence of FSRQs and BL Lacs in radio and
X-ray selected samples respectively, the higher redshifts and evolution of
FSRQs as compared to BL Lacs, and the no evolution of X-ray selected BL
Lacs, were confirmed, which shows that they are independent of the details
of the LF and evolution.

We also modified our assumption that standard accretion disks are
associated only with sources having an FR II parent but the simulations we
obtained were inconsistent with observations. In fact, even if the
relatively small fraction of $20\%$ of objects with FR I parents are
randomly assigned an accretion disk and allowed to evolve, X-ray selected
BL Lacs reach too high values of $\langle V/V_{\rm m} \rangle$ and $\langle
z \rangle$, 0.58 and 0.91 respectively, which are incompatible with
observations.

\section{Discussion}\label{discussion}

So far, blazars have been largely discovered in the radio, $\mu$-wave, X-ray
or \gr bands.  We have shown that the wide diversity of \nup~observed in
blazars results in strong differences in the observed flux at radio or
X-ray frequencies (up to three orders of magnitudes) causing severe
selection effects in surveys performed in these energy bands.  This, combined with
the fact that blazars are always identified on the basis of their optical
spectrum, which reflects a mix of non-thermal, thermal, and host galaxy
emission, has been the origin of a number of enduring open issues about the nature
of blazars.

\subsection{Evolution}

A long-standing blazar puzzle is the difference in redshift distribution
and cosmological evolution between FSRQs and BL Lacs, selected both in the
radio and in the X-ray band. BL Lacs are mostly located at low redshifts
and exhibit moderate, or even negative, evolution, while FSRQs evolve
strongly just like radio quiet QSOs and show a redshift distribution that
peaks at $z > 1$ \citep{stocke82,stickel1991,rec00,beck03,pad07,GiommiWMAP09}.
Our simulations reproduce quite well both of these findings (see
Fig. \ref{fig:wmap_red} and Tables \ref{tab:radiosim} and \ref{tab:xsim})
implying that they are due to heavy selection effects.  Most of the
simulated BL Lacs found in radio surveys are luminous objects with broad
lines that are diluted by non-thermal radiation beyond the 5~\AA~EW limit
(many of them just below, thus allowing a measurement of their redshift),
while the BL Lacs found in simulated X-ray surveys typically show high
\nup~values (and therefore are X-ray bright) and are drawn from the
low-power end of the radio luminosity function where non-evolving FR Is are
preferentially found (Section \ref{ingredients}). We note that the
\avvovm~of radio selected BL Lacs is not too different from that of FSRQs,
while the \avvovm~of X-ray selected BL Lacs is significantly lower, as
found in real surveys (Section \ref{radio_survey}).

Another interesting outcome of our simulations is the fact that X-ray selected BL
Lacs with progressively larger values of \nup~are characterized by lower
and lower values of \avvovm~(see Tab. \ref{tab:xsim}).  
This is in full agreement with the puzzling, and so far unexplained, results 
of  \cite{rec00} and of \cite{giommi_sedent_I} who reported that the \avvovm~of 
BL Lacs is a function of their X-ray-to-radio flux ratio.

\subsection{\nup~distribution}

Recent results, based on radio and $\gamma$-ray surveys, have revealed that
BL Lacs, on average, display a distribution of \nup~energies, which is
shifted to values higher than those of FSRQs \citep{abdosed,GiommiPlanck},
expanding on the well-known fact that high \nup~objects (HSPs) are always
BL Lacs.  This experimental difference is well reproduced in our
simulations (see Fig. \ref {fig:radio_nupeak}) which give $\langle
$log(\nup)$\rangle$ = 12.9 for FSRQs and $\langle $log(\nup)$\rangle$ =
14.1 for BL Lacs for the case of a radio survey. This distinction is due to the fact that
blazars with higher \nup~values produce more non-thermal optical light than
low \nup~sources, diluting more easily the broad line component, and are
therefore classified more frequently as BL Lac objects. We note that this has been
interpreted in the literature as an intrinsic physical difference between
LSP (detected mostly in the radio band) and HSP (detected mostly in the
X-ray and \gr band) blazars due to the fact that HSPs are observationally
characterized by a low intrinsic power and external radiation field, given
their very weak or absent emission lines. As a consequence, cooling was
thought to be less dramatic in HSP allowing particles to reach energies
high enough to produce synchrotron emission well into the X-ray band
\citep{ghis98}. In our scenario, instead, all sources have exactly the same
chance of being HSP or LSP (that is, the value of $\gamma_{\rm peak}$ is
drawn from the distribution shown in Fig. \ref{fig:gammae} {\it
  independently} of luminosity) and the very different \nup~distributions
observed in radio and X-ray surveys arise from the strong selection effect discussed in Section \ref{sec:seq} and the emission line
dilution mentioned above.

Our simulations predict the existence of a significant number of BL Lacs
with redshift that cannot be measured, which occurs when both \nup~and
radio power are so large that dilution becomes extreme.  For example, $\sim
81\%$ of our simulated sources with \nup~$> 10^{15}$ Hz and $P_{\rm r} >
10^{26}$ W/Hz have no redshift.  This is consistent with the fact that most
BL Lacs in current \gr~selected samples have no measured redshift, as
\fermi is known to preferentially select high \nup~BL Lacs
\citep{fermi1lac,fermi2lac}.  This effect is also clearly shown in
Fig. \ref{fig:sequence} where most of the simulated blazars with no
measurable redshift (filled circles with light colors) occupy the top right
part of the diagram. 
Fig. \ref{fig:zdistOfnoredshiftSources} shows the intrinsic redshift distribution of these featureless BL Lacs 
for the case of our simulation of a radio flux density limited survey.

\subsection{What is a BL Lac?}

\cite{bla78} had originally suggested that the absence of broad lines in BL
Lacs was due to a very bright, Doppler-boosted synchrotron continuum. In
the years following that paper observations of various BL Lacs, mostly
selected in the X-ray band, showed that in many cases their optical
spectrum was not swamped by a non-thermal component, as host galaxy features
were very visible, and it was thought that most BL Lacs had intrinsically
weak lines. We have shown here that these two possibilities are not
mutually exclusive and indeed are both viable, depending on radio power
and, therefore, on the band of selection. One important consequence of our
scenario is that objects so far classified as BL Lacs on the basis of their
{\it observed} weak, or undetectable, emission lines belong to two
physically different classes: intrinsically weak lined objects, whose
parents are LERGs/FR Is (more common in X-ray selected samples, since they
reach lower radio powers) and heavily diluted broad-lined sources, which
are beamed HERGs/FR IIs (more frequent in radio selected
samples). Therefore, while the non-thermal engine is probably the same, the
thermal one is obviously different. This solves at once the issue of the
FSRQ/BL Lac transition objects and of the many differences between BL Lacs
selected in the radio and X-ray bands, which include line strength,
extended radio emission and morphology, and evolution \citep[e.g.][and
  references therein]{rec01}.  This hypothesis, which explains also many
other open issues of blazar research, is directly testable. Our
simulations, in fact, imply that the majority of sources in high-flux
density radio-selected samples are identified as BL Lacs only because all
lines have EW~$<~5$~\AA~{\it in the optical observing window}. These
sources should show a strong (EW $> 5$ \AA) H$\alpha$ line if observed in
the near or mid-infrared, since this is the strongest emission line,
 and would then be considered FSRQs, which
reinforces our point on the very strong effect that line dilution has on
blazar classification. HERG/FR II BL Lacs, then, should be easily
recognizable through infrared spectroscopy. These sources, as expected, are
also more dominant at higher radio powers: for example, according to our
radio simulation the fraction of BL Lacs with standard accretion disks is
only $\sim 3\%$ for $P_{\rm r} \le 10^{26}$ W/Hz but becomes $\sim 98\%$
above this value. This also means that fainter radio-selected samples of BL
Lacs should be more and more similar to X-ray selected ones, apart from
their \nup~values, since higher \nup~are easier to detect in the X-rays
\citep{padgio95}.

The implications of our hypothesis for unified schemes is quite
straightforward: the parent population of BL Lacs need to include both
LERGs/FR Is and HERGs/FR IIs, while that of FSRQs is made up of HERGs/FR
IIs only. This should have only a small effect, for example, on the LF
fitting done by \cite{UP95}, as HERG/FR II BL Lacs make up the high-power
end of the radio LF while most of the number density comes from LERG/FR I
BL Lacs.

\subsection{Blazar classification}

Our new scenario has strong implications on blazar classification as well.
If the relevant physical distinction for radio sources is between LERGs
(mostly FR Is) and HERGs (FR IIs), for most purposes then (LFs, evolution,
etc.) HERG/FR II BL Lacs should be simply grouped with FSRQs. How does one
distinguish in practice HERG/FR II BL Lacs from LERG/FR I BL Lacs?  This is
simple in the presence of {\it any} (even weak) broad lines or for
transition objects. In other cases (e.g. completely featureless spectrum or 
in presence of absorption features) there is no easy way to distinguish
between the two subclasses although, for example, radio power and/or
morphology could help. Given the paucity of known LERG/FR Is at relative
high redshifts, X-ray selected (and also fainter radio-selected) BL Lac
samples are also useful in selecting such sources, which are very relevant
also for the study of the so-called ``AGN feedback'' and the role that AGN
radio emission plays in galaxy evolution through the so-called
``radio-mode'' accretion \citep{cro06}. It should also be clear that BL
Lacs can be used to study the broader issue of the relationship between
LERGs and HERGs, including their evolution.
    
A by-product of our simulations has also been the realization that some
sources classified as radio-galaxies do {\it not} have their jets oriented
at large angles with respect to the line of sight, as expected, but are
instead moderately beamed blazars with their non-thermal emission swamped
by the galaxy (note that none of these objects has a standard accretion
disk). These sources, which are all local ($\sim 90\%$ at $z \le 0.07$)
should be recognizable by their blazar-like SEDs and indeed some of them
have already been identified by \cite{den00}, \cite{gio02,giommi_sedent_II} and
\cite{ant05}.
 
Recently \cite{ghis11} have proposed a new classification scheme to divide BL Lacs
from FSRQs, which is based on the broad line region (BLR) luminosity in
Eddington units and set at a dividing value of $L_{\rm BLR}/L_{\rm Edd}
\sim 5 \times 10^{-4}$. This turns out to be also the value, which
separates radiatively efficient (i.e., standard accretion disks) from
radiatively inefficient (i.e., ADAFs) regimes, and therefore coincides with
our HERG/FR II -- LERG/FR I division. Therefore, \cite{ghis11} are also
suggesting that HERG/FR II BL Lacs belong with the FSRQs.

\section{Conclusions}

We have tackled the open issue of the relationship between the various
blazar subclasses, and of the consequences of selection effects, through
extensive Monte Carlo simulations. Our approach is based on robust observational input,
and on various results obtained by many authors over the past 20
years or so, which had never before been put together in a comprehensive
way. We kept the number of assumptions and parameters
to a minimum so as to draw robust conclusions.  Our starting point are two
populations of high-excitation, high radio-power, evolving and
low-excitation, low-power, non-evolving radio sources. To these we add: an
homogeneous synchrotron self-Compton component with a distribution of
electron Lorentz factors peaked at relatively low values, an optical/UV
quasar template, the EW distributions of the main broad lines observed in
radio-quiet quasars, a distribution of disk/jet ratios, and finally an
elliptical host galaxy.

Our main results can be summarized as follows:

\begin{enumerate} 
\renewcommand{\theenumi}{(\arabic{enumi})} 
 
\item we explain the main properties of all blazar subclasses, namely FSRQs
  and BL Lacs selected in the radio and X-ray band.  Our simulations
  reproduce well the prevalence of FSRQs and BL Lacs in radio and X-ray
  selected samples respectively, the higher redshifts, evolution and
  lower \nup~of FSRQs as compared to BL Lacs, the non evolution of X-ray
  selected BL Lacs, and the main differences between radio and
  X-ray-selected BL Lacs in a manner that does not dependent significantly 
  on the adopted LF and evolutionary details;

\item objects classified until now as BL Lacs on the basis of their {\it
  observed} weak, or undetectable, emission lines actually belong to two
  physically different classes: most BL Lacs in high flux density,
  radio-selected samples are actually beamed radio quasars with their
  emission lines heavily diluted by the non-thermal continuum and for all
  purposes should then be grouped with FSRQs. Infrared spectroscopy should
  reveal in many such sources the presence of strong (EW $> 5$~\AA)
  H$\alpha$; most BL Lacs selected in the X-ray band are instead
  intrinsically weak-lined, low-excitation radio galaxies with a strong
  non-thermal, jet component;
 
\item there are only two main intrinsic blazar types: low-ionization
  (mostly beamed FR Is) and high-ionization (beamed FR IIs) ones. All other
  classifications and classes proposed so far are not physically relevant
  and are simply due to severe selection effects and different relative
  strengths of the non-thermal, thermal, and galaxy components, which make
  up blazars' optical emission;
   
\item some sources classified as radio-galaxies on the basis of their
  optical properties are instead moderately beamed blazars (that is, sources with their jets
  forming a small angle with respect to the line of sight) with their
  non-thermal emission swamped by the host galaxy. Some of these sources have
  already been recognized from their SEDs;
 
\item our simulations show that the purported correlation between
  luminosity and \nup, the so-called ``blazar sequence'', is likely a
  selection effect resulting from comparing shallow radio surveys with
  shallow X-ray surveys.  We also show that blazars with featureless
  optical spectra, and therefore without a redshift determination, are
  mostly high luminosity -- high \nup~sources, a type of blazar that should
  not exist in the ``blazar sequence'' scenario. This is consistent with
  the lack of redshift in the majority of BL Lacs in current \gr selected
  samples, as \fermi is known to preferentially select high \nup~BL Lacs;
  
\item the topics addressed in this paper are not relevant only to blazars
  but are also related to the broader issues of low-ionization radio
  galaxies and radio-mode ``AGN feedback''.

\end{enumerate}

In this paper we limited our simulations to the radio and the X-ray
bands where SSC is a fair approximation of the observed non-thermal
emission. The properties of \gr~detected blazars are instead not consistent with
simple SSC models \citep[e.g.][]{abdosed}, and almost half of the radio and X-ray selected 
LSP blazars are  \gr quiet  \citep[e.g.][]{GiommiPlanck}. 
The present approach must therefore be integrated with additional information about the properties 
of the inverse Compton emission before it can be used to simulate \gr surveys. 
We are planning to extend our simulations to the \gr band by taking into account the recent results
of \citet{GiommiPlanck} who determined the \gr properties of blazar samples
selected in different bands.

\section*{Acknowledgments}

We thank A. Pollock for careful reading of the manuscript and useful suggestions.
ST would like to thank Mark Subbarao and Rich Plotkin for useful
discussions on the SDSS spectroscopic pipeline. We also thank Matteo Perri
for providing part of the software used for our Monte Carlo simulations. We
acknowledge the use of data and software facilities from the ASI Science
Data Center (ASDC), managed by the Italian Space Agency (ASI). Part of this
work is based on archival data and on bibliographic information obtained
from the NASA/IPAC Extragalactic Database (NED) and from the Astrophysics
Data System (ADS). Funding for the SDSS and SDSS-II was provided by the
Alfred P. Sloan Foundation, the Participating Institutions, the National
Science Foundation, the U.S. Department of Energy, the National Aeronautics
and Space Administration, the Japanese Monbukagakusho, the Max Planck
Society, and the Higher Education Funding Council for England. The SDSS was
managed by the Astrophysical Research Consortium for the Participating
Institutions.


\label{lastpage}

\end{document}